\documentclass[structabstract]{aa}  
\usepackage{graphicx}
\usepackage{txfonts}
\usepackage{natbib}
\newcommand{\teff}  {T$_\mathrm{eff}$}
\newcommand{\logg}  {$\log g$}
\newcommand{\kms}{\hbox{${\rm km}\:{\rm s}^{-1}$}}

\def\aap{A\&A}

\def\apjs{ApJS}
\def\apjl{ApJL}

\begin{document}
   \title{Li abundances in F stars: planets, rotation and galactic evolution\thanks{
Based on observations collected at the La Silla Observatory, ESO
(Chile), with the HARPS spectrograph at the 3.6 m ESO telescope, 
with CORALIE spectrograph at the 1.2 m Euler Swiss telescope and 
with the FEROS spectrograph at the 1.52 m ESO telescope;
at the Paranal Observatory, ESO (Chile), using the UVES spectrograph
at the VLT/UT2 Kueyen telescope, and with the FIES and SARG
spectrographs at the 2.5 m NOT and the 3.6 m TNG, respectively, 
both at La Palma (Canary Islands, Spain).
}}

   \author{E. Delgado Mena\inst{1,2}
      \and S. Bertr\'an de Lis\inst{3,4} 
      \and V. Zh. Adibekyan\inst{1,2}
      \and S. G. Sousa\inst{1,2}
      \and P. Figueira\inst{1,2}
      \and A. Mortier\inst{6}
      \and J.~I.~Gonz\'alez Hern\'andez\inst{3,4}
      \and M. Tsantaki\inst{1,2,3}
      \and G. Israelian\inst{3,4} 
      \and N.~C.~Santos\inst{1,2,5}
      }
\institute{
Centro de Astrofisica, Universidade do Porto, Rua das Estrelas,
4150-762, Porto, Portugal 
             \email{Elisa.Delgado@astro.up.pt}
\and
Instituto de Astrof\'isica e Ci\^encias do Espa\c{c}o, Universidade do Porto, CAUP, Rua das
Estrelas, PT4150-762 Porto, Portugal
\and 
Instituto de Astrof\'{\i}sica de Canarias,
C/ Via Lactea, s/n, 38200, La Laguna, 
Tenerife, Spain 
\and 
Departamento de Astrof\'isica, Universidad de La Laguna, 38205 La Laguna, Tenerife, Spain
\and
Departamento de F\'isica e Astronom\'ia, Faculdade de Ci\^encias, Universidade do Porto, Portugal
\and
SUPA, School of Physics and Astronomy, University of St Andrews, St Andrews KY16 9SS, UK
}     


   \date{Received ...; accepted ...}

 
  \abstract
{}
{The goal of this work is, on the one hand, to study the possible differences of Li abundances between planet hosts and stars without detected planets at effective temperatures hotter than the Sun, and on the other hand, to explore the Li dip and the evolution of Li at high metallicities.}
{We present lithium abundances for 353 Main Sequence stars with and without planets in the \teff\ range 5900-7200 K. 265 stars of our sample were observed with HARPS spectrograph during different planets search programs. The remaining targets have been observed with a variety of high resolution spectrographs. The abundances are derived by a standard LTE analysis using spectral synthesis with the code MOOG and a grid of Kurucz ATLAS9 atmospheres.}
{We find that hot jupiter host stars within the \teff\ range 5900-6300K show lower Li abundances, by 0.14 dex, than stars without detected planets. This offset has a significance at the level 7$\sigma$, pointing to a stronger effect of planet formation on Li abundances when the planets are more massive and migrate close to the star. However, we also find that the average v \textit{sin}i of (a fraction of) stars with hot jupiters is higher on average than for single stars in the same Teff region suggesting that rotational-induced mixing (and not the presence of planets) might be the cause for a greater depletion of Li. We confirm that the mass-metallicity dependence of the Li dip is extended towards [Fe/H] $\sim$ 0.4 dex (beginning at [Fe/H] $\sim$ -0.4 dex for our stars) and that probably reflects the mass-metallicity correlation of stars of the same \teff\ on the Main Sequence. We find that for the youngest stars ($<$ 1.5 Gyr) around the Li dip, the depletion of Li increases with v \textit{sin}i values, as proposed by rotationally-induced depletion models. This suggests that the Li dip consists of fast rotators at young ages whereas the most Li-depleted old stars show lower rotation rates (probably caused by the spin-down during their long lifes). We have also explored the Li evolution with [Fe/H] taking advantage of the metal-rich stars included in our sample. We find that Li abundance reaches its maximum around solar metallicity but decreases in the most metal-rich stars, as predicted by some models of Li Galactic production.}
{}


   \maketitle
%

\section{Introduction}

Lithium is one of the most studied chemical elements in the literature. Despite all the efforts done to unveil the mechanisms of production and destruction of this interesting element, there are still some unsolved misteries. For instance, the disagreement found between the abundance of the most metal-poor stars in the Galaxy (the so-called 'Spite plateau' with A(Li)\footnote{${\rm A(Li)}=\log[N({\rm Li})/N({\rm H})]+12$}$\sim$2.2, \citet{spite82}) and the initial primordial abundance given by the WMAP observations (A(Li)$\sim$2.7, \citet{steigman10,cyburt08}) is not understood yet. Moreover, the current Galactic Li production models \citep[e.g.][]{prantzos12} are not able to yield enough Li to explain the meteoritic abundance of 3.31 \citep{anders} or the maximum Li abundances found in young clusters \citep[e.g.][]{sestito05}. On the other hand the standard model of Li depletion (which only considers convection), \citep[e.g.][]{deliyannis90,pinsonneault97}, cannot explain the observed Li abundances in solar-type stars or in mid-F stars that have undergone the Li dip. Furthermore, in the last years a new discussion about the effect of planets on the depletion of Li has been opened \citep[e.g.][hereafter DM14]{israelian09,ramirez_li12,gonzalez_li14,figueira14,delgado14}.\\

Lithium, as other light elements, can be easily destroyed in stellar interiors by p-$\alpha$ reactions. Although Li depletion occurs primarily in the pre-Main Sequence (PMS), it can also take place in stellar envelopes if any extra mixing process exists. Therefore, its abundance can provide us important information about the internal structure of stars. In this work we present homogeneous Li abundances for a sample of 353 'hot' stars (early G and F stars) with a wide range in metallicities and ages. We exploit the metal-rich stars in our sample to study the behaviour of the Li dip and the chemical evolution of Li at high metallicities. Finally, we also investigate if the presence of planets affect Li abundances for these hotter stars. \\

This paper is divided as follows: Section 2 briefly describes the collected data for this work together with the determination of stellar parameters and abundances of lithium. In Section 3 we discuss the results related to different topics: the connection of Li abundances with the presence of planets; the behaviour of the Li dip; the chemical evolution of Li at high metallicities and the Li distribution in the galactic disks. We finalize with the summary in Section 4.

\begin{figure}
\centering
\includegraphics[width=9.2cm]{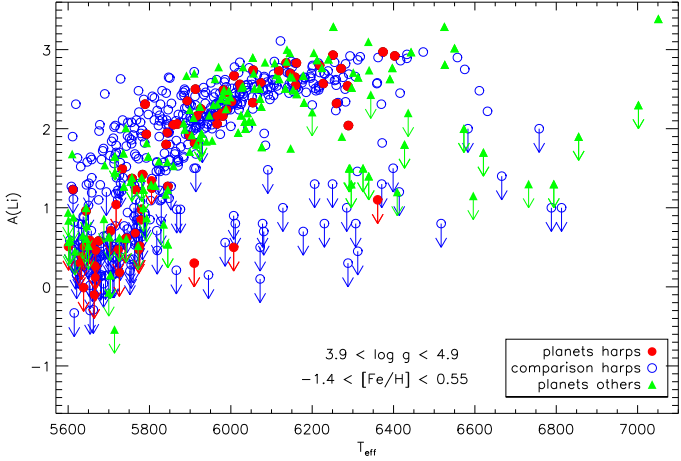}
\caption{Lithium abundances vs. \teff ~for planet host stars (red filled circles) and comparison sample stars (blue open circles) from HARPS together with other planet hosts (green triangles). Down arrows represent A(Li) upper limits.} 
\label{Liteff_todas}
\end{figure}

\section{Observations and analysis}

The baseline sample used in this work is formed by 1111 FGK stars observed within the context of the HARPS GTO programs. It is a combination of three HARPS sub-samples hereafter called HARPS-1 \citep{mayor03}, HARPS-2 \citep{locurto} and HARPS-4 \citep{santos_harps4}. The individual spectra of each star were reduced using the HARPS pipeline and then combined with IRAF\footnote{IRAF is distributed by National Optical Astronomy Observatories, operated by the Association of Universities for Research in Astronomy, Inc., under contract with the National Science Fundation, USA.} after correcting for its radial velocity shift. The final spectra have a resolution of R $\sim$110000 and high signal-to-noise ratio (55\% of the spectra have S/N higher than 200). The total sample is composed by 135 stars with planets and 976 stars without detected planets. For this work, we mainly focus on the hottest \teff\ ($>$5900 K) where we have 36 and 229 stars with and without planets, respectively. All the planet hosts and non-hosts stars are listed in Tables~\ref{tabla_harps_plan} and 4, respectively.
To increase the number of stars with planets we used high resolution spectra for 88 planet hosts (see Table 5) which come from different observing runs and spectrographs. We refer the reader to Table 1 of DM14 for a detailed list of those instruments. The data reduction was made using the IRAF package or the respective telescopes pipelines. All the images were flat-field corrected, sky substracted and co-added to obtain 1D spectra. Doppler correction was also done.\\

The stellar atmospheric parameters were taken from \citet{sousa08,sousa_harps4,sousa_harps2} for the HARPS samples and from \citet{santos04, santos05, sousa06, mortier13_transits, santos13} for the rest of the planet hosts. All the sets of parameters were determined in a homogeneous way. Lithium abundances ${\rm A(Li)}$, stellar masses and ages were derived in the same way as DM14. We refer the reader to that work for further details about the determination of stellar parameters and Li abundances.\\

The determination of rotational projected velocities (v \textit{sin}i values) was done with a combined Fourier transform and goodness-of-fit methodology using the IACOB program \citep{simondiaz14}. We could only determine it for the stars with spectra of S/N ratios above $\sim$100.\\

\begin{figure}
\centering
\includegraphics[width=9.0cm]{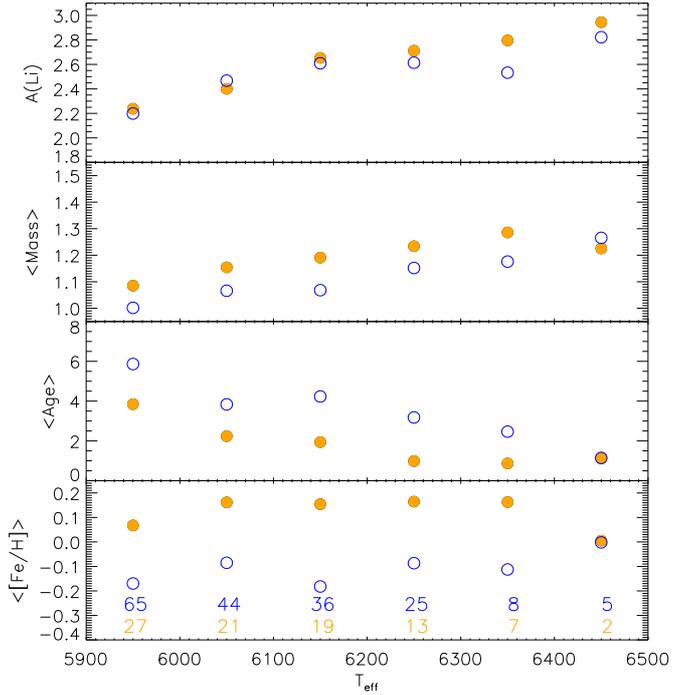}
\caption{Average of A(Li), stellar mass, age and [Fe/H] in 6 \teff~ bins for all (HARPS+others) the planet host stars (orange filled circles) and comparison sample stars (blue open circles). The number of stars in each bin is indicated with the respective color. Only stars with Li detections are considered.} 
\label{averages}
\end{figure}

\begin{table*}
\label{test}
\centering
\begin{tabular}{cccccccccc} \hline\hline
sample of planet hosts & number & {\it int.} &$\beta_1$ & $\beta_2$  & $\beta_3$  & $\beta_4$ & $\beta_5$  & offset & significance \\ 
\hline
jupiter size hosts & 87PH \& 176CS & -64.04 & 17.81 & 0.02 & -0.18 & -0.21 & & -0.07 & 6.5$\sigma$ \\
hot jupiter hosts & 24PH\& 176CS & -69.64 & 19.37 & -0.05 &-0.24 &  -0.28 & & -0.14 & 7.0$\sigma$ \\
\hline 
\hline
analysis including v \textit{sin}i & & & & & & \\
\hline 
jupiter size hosts & 47PH \& 62CS &  -86.39 & 23.72 & 0.06 & -0.10 &  -0.20 & -0.12 & 0.04 & 2.6$\sigma$ \\
hot jupiter hosts & 11PH \& 62CS& -98.37 & 27.16 & -0.05 & -0.29 & -0.28 & -0.15 & 0.08 & 2.1$\sigma$ \\ 
\hline 
\hline
\end{tabular}
\caption{The parameters for each coefficient as resulting from Multivariable Linear Regression Analysis in the four tests (PH are planet hosts and CS are the comparison stars). The offset is only included in the fit for planet hosts since for comparison stars is 0 by definition. The last column reflects the siginificance of the offset found between both smaples.}
\end{table*}

\section{Results and discussion}

\subsection{General behaviour of Li in F stars}\label{li-general}
 
In Fig.~\ref{Liteff_todas} we present a general overview of the Li abundances as a function of effective temperature for our sample. The ranges in [Fe/H] and gravity for the stars in this sample are specified in the plot. In order to better appreciate the behaviour of Li in a wider \teff\ range we have also included the solar type stars from DM14, with 5600 K $<$ \teff\ $<$ 5900 K.
As expected, Li abundances decrease as the stars get cooler due to their thicker convective envelopes.
However, we can still observe an important number of stars with a strong destruction of Li. The stars around ~6400 K belong to the well known Li dip, first discovered in the Hyades cluster by \citet{boesgaard86} but those with cooler temperatures, between 5900K and 6200K, are not so common in studies of clusters or field stars. We would expect these stars to have higher Li abundances unless they are evolved stars from the dip, as suggested by \citet{chen01}. These objects will be further studied in a separate work. Although we do not expect to have evolved stars in our sample we have removed the stars with \logg $<$ 4.2 since our spectroscopic \logg\ values could be overestimated for the hotter stars \citep{mortier14}.

\subsection{Li and planets}
The Li dependence on the presence of planets has been extensively discussed in the literature. On the one hand several independent groups find that planet hosts with \teff\ close to solar present lower abundances of Li when compared to non-hosts \citep{israelian04,takeda05,chen06,gonzalez_li08,israelian09,takeda10,gonzalez_li10,sousa_li,delgado14,gonzalez_li14,figueira14}. On the other hand other authors do not find such a dependence \citep{ryan00,luck06,baumann,ghezzi_li,ramirez_li12}.
\citet{gonzalez_li08} proposed that stars with planets around 6100K show higher Li abundances than stars without detected planets. However, after increasing the sample size, the same author discarded this effect and presented weak evidence that planet hosts at \teff\ $\sim$ 6100-6200K are deficient in Li compared to stars without detected planets \citep{gonzalez_li14,gonzalez_li15}. 
Visually we cannot pinpoint any strong difference in the Li abundance detections between stars with and without planets in Fig.~\ref{Liteff_todas}. However it is quite clear that in the \teff\ range between 5900K and 6300K there are relatively more non-hosts with upper limits in Li abundances. This feature was also pointed out by \citet{ramirez_li12}.\\

In Fig. \ref{averages} we compile the average values of Li abundance detections and other parameters for stars with and without detected planets (in bins of 100 K). Since Li abundances depend on several parameters (e.g. \teff, [Fe/H], age) one should be cautious when comparing stars and construct samples the least biased possible (for a further discussion see DM14).
For example, in all these subsamples except the hottest one, planet hosts are younger and also more metal-rich on average, as expected \citep[e.g.][]{santos04}. Nevertheless, this difference in parameters does not seem to affect too much the degree in Li depletion (see Section \ref{li-feh}) except maybe in the \teff\ range 6300-6400K where we observe the highest difference in Li. For the rest of subamples the average values of stars with and without planets are quite similar and within the errors.\\ 

In order to remove the effect of different stellar parameters when comparing Li abundances we apply a multivariate regression fit to the planet host sample and the comparison sample as done in \citet{figueira14}: 
\vskip-1.5em
\begin{eqnarray}\label{eq1}
\mathrm{log(A(Li))} & = & int. \,+\, \beta_1 \mathrm{log(}\rm T_{\rm eff}\mathrm{)} \,+\, \beta_2 \rm [Fe/H] \,+\, \beta_3 \mathrm{log}\,g \,+ \\ \nonumber
                       && + \, \beta_4 \mathrm{log(Age)} \,+\, M\,\times\,\mathrm{offset}
\end{eqnarray}

\begin{figure*}
\centering
\includegraphics[width=18.0cm]{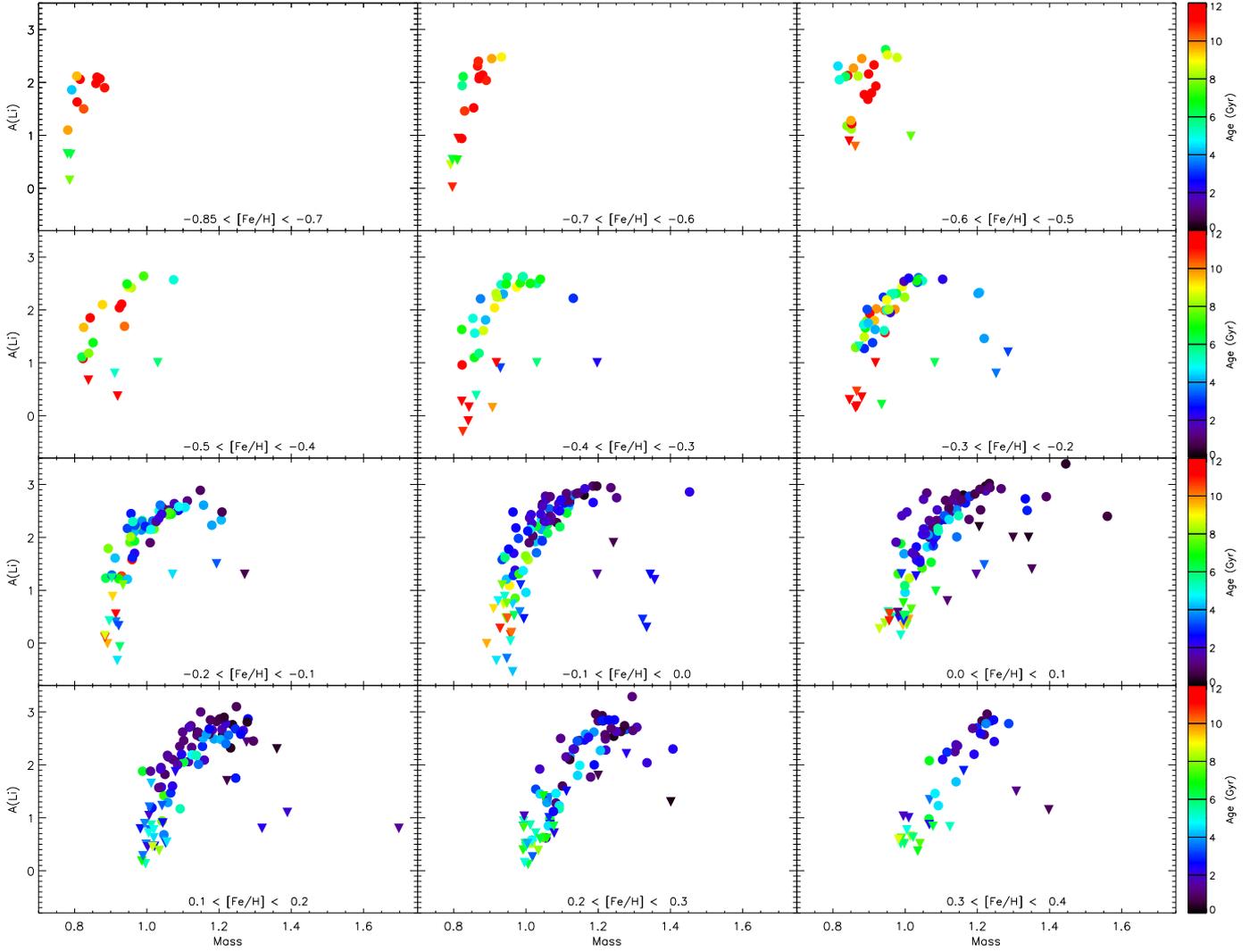}
\caption{Lithium abundances vs. stellar mass in twelve metallicity bins for all our stars with \logg $>$ 4.2. Downward triangles represent A(Li) upper limits. The ages are depicted by a colour scale.} 
\label{Li-masa}
\end{figure*}

On both samples the same linear dependence of Li on stellar parameters is assumed but allowing an offset for the planet host sample, that is, \textit{M}=0 for the comparison sample and \textit{M}=1 for the planet host sample. By doing so, we ensure that a possible difference in Li abundance is not due to different stellar parameters. For this calculation we consider all our stars with 5900 K $<$ \teff\ $<$ 6300 K and Li detections (we exclude the upper limits), 87 planet hosts (with Jupiter type planets: M$_{P}$ $\geq$ 0.1 M$_{J}$) and 176 comparison stars. We chose to cut at \teff~ = 6300K to allow a fair comparison with \citet{gonzalez_li15} sample. Moreover, that is roughly the temperature at which the Li dip begins to develop and it would be difficult to distinguish a possible effect of planets on Li abundances from other depletion mechanisms.\\

The results are shown in Table 1. As expected, the strongest dependence lies on \teff, while it is very small for the other parameters. We find that the planet host sample shows a depletion of 0.07 dex with respect to non-hosts. Although this offset is significant (at 6.5$\sigma$ level), its significance is naturally heavily dependent on the error bars. If we artificially increase the error bars by a factor of 2 or 3 the significance drops to 3.3$\sigma$ and 2.2$\sigma$, respectively. This offset agrees with the results by \citet{gonzalez_li15} though it is quite small and at the level of the uncertainties\footnotetext{The average uncertainty in Li abundances for our stars is 0.07 dex}. Therefore, it seems that the effect of giant planets observed for solar analogues is not obvious for hotter stars, probably due to their shallower convective envelopes. As explained in DM14, the effect of planets on Li abundances is expected to be higher for more massive planets (stronger effect on rotational history, \citet{bouvier08} and for planets that migrate (more violent accretion bursts, \citet{baraffe10}). Thus, we explore the behaviour of stars hosting hot jupiters since these planets are massive and some theories of planet formation predict a migration close to the star \citep[e.g.][]{alibert05}. Then, we repeat our previous calculation but using as planet hosts sample those stars which hosts planets with M $>$ 0.1M$_{J}$ and P $<$ 5 days. In this case we find a higher offset than before, -0.14 dex, with a significance level of 7$\sigma$. As before, we increase the error bars by a factor of 2 and 3, which drops the significance to 3.6$\sigma$ and 2.4$\sigma$, respectively. We note that our sample of hot jupiter hosts is small (24 stars with 5900 K $<$ \teff\ $<$ 6300 K) but this is an interesting result that deserves to be explored further in the future.\\

Finally, we investigate the possible effect of rotation on Li abundances for our sample of planet hosts. The models of rotationally-induced mixing predict that during the Main Sequence (MS), stars with higher rotation rates are expected to deplete more Li than slower rotators. We find that the average v \textit{sin}i for hot jupiter hosts in this \teff\ range is larger, 5 \kms\ (derived only for 11 stars) than for the comparison stars, 3.1 \kms (derived for 62 stars), hence this could explain the offset previously found for stars with hot jupiters. In order to test this effect we repeat the same analysis as before but including the v \textit{sin}i in the equation and forcing a same dependence on it both for the planet host sample and in the comparison stars sample:
\vskip-1.5em
\begin{eqnarray}\label{eq2}
\mathrm{log(A(Li))} & = & int. \,+\, \beta_1 \mathrm{log(}T_{eff}\mathrm{)} \,+\, \beta_2 [Fe/H] \,+\, \beta_3 \mathrm{log}\,g \,+ \\ \nonumber
                       && + \, \beta_4 \mathrm{log(Age)} \,+ \, \beta_5 \mathrm{v \textit{sin}i} \,+\, M\,\times\,\mathrm{offset}
\end{eqnarray}

The results are shown in the second part of Table 1. As expected from the models of rotationally-induced mixing, Li abundances show a negative dependence on v \textit{sin}i. The  offsets are now positive (i.e. higher Li abundances for planet hosts) but they are also quite less significant than before ($2.6\sigma$ and $2.1\sigma$, for the jupiter size planets and the hot jupiters, respectively). This result points to an effect of rotation on Li abundances though we have to be cautious since our sample of measured v \textit{sin}i values is quite small and potentially biased (we could only measure v \textit{sin}i in 42\% of our stars with Li detections in this \teff\ range).

\subsection{The Li dip: dependence on metallicity, age and v \textit{sin}i}

The Li dip was first discovered in the Hyades by \citet{boesgaard86}. For clusters younger than the Pleaides ($\sim$200 Myr) this feature is not observed and stars more massive than a solar mass show a constant maximum value of A(Li)=3-3.2 dex \citep{lambert04}. Therefore, the Li dip has to be formed during the MS. Indeed, the maximum Li abundance is similar for the youngest clusters than for the slightly older ones (300 Myr-2Gyr), hence, F stars in the MS experience very little depletion up to ages of 1 Gyr. For older clusters there are hints of the presence of the dip though not always there is a significant number of stars at those temperatures. The Li dip is not well defined in Fig.~\ref{Liteff_todas} because in our sample there are stars of different ages and metallicities. In order to appreciate a clearly shaped Li dip one should divide the stars by ranges with similar [Fe/H] and age, as happens in open clusters. \\
 
In Fig. \ref{Li-masa} we show Li abundances as a function of mass in 12 different metallicity bins for the stars in this work together with the sample of DM14 (we exclude the most metal-poor and most metal-rich bins due to their low number of stars). From this plot we can confirm that the Li dip happens at higher masses as the metallicity increases. This fact was first suggested by \citet{balachandran95} who found that the mass at which the dip occurs depends on the stellar metallicity, while the ZAMS (Zero Age Main Sequence) \teff\ does not. Later studies on clusters \citep[e.g.][]{cummings12,francois13} or in field stars \citep[e.g.][]{lambert04} have confirmed this feature. For example in the [Fe/H] range [-0.6,-0.5] we have a unique star at the Li dip with a mass of 1.02 M$_{\odot}$. This agrees with the Li dip center of 1.06M$_{\odot}$ found by \citep{francois13} in a similar metallicity cluster, NGC2243 with [Fe/H] = -0.54 dex. On the other hand, at the higher metallicities we find two stars with 1.3 and 1.4 M$_{\odot}$ in the Li dip, which compares well with the mass of the cool side of Li dip in NGC6253 (1.34 M$_{\odot}$, [Fe/H] = 0.43 dex, \citet{cummings12}). In that work they also compare their results with the Hyades which has a Li dip mass of 1.27 M$_{\odot}$. In our field stars of similar metallicity (0.1-0.2dex) the Li dip seems to be at $\sim$1.3 M$_{\odot}$. From Fig. \ref{Li-masa} we can see that the increase of mass with metallicity not only happens for the Li dip stars but for all the objects within each metallicity bin, therefore this is just a reflection of the mass-metallicity correlation for stars with similar \teff\ in the MS and confirms the suggestion by \citet{balachandran95}.\\

In Fig. \ref{Li-masa} we also show the ages of the stars by a color scale. For the more metal-poor bins we cannot observe the Li dip because our MS stars are too old and thus too cool to be susceptible to that process. For instance, for older clusters like M67 the Li dip is formed by subgiants. We have cleaned our sample of possible evolved stars so the hotter stars that usually form the Li dip have to be young enough and not evolved yet. In fact, if we observe the Li dip at different metallicities it is always formed by stars younger than $\sim$4 Gyr, with the age slightly decreasing as the metallicity increases.\\

\begin{table*}
 \caption{Li abundances for the fast rotators of section 3.3. Stellar parameters from \citet{tsantaki14}.}
\centering
\begin{tabular}{lcccrrcrc}
\hline
\hline
\noalign{\smallskip}
Star & $T_\mathrm{eff}$ & $\log{g}$ & [Fe/H] & Age & Mass & vsini & ${\rm A(Li)}$ \\  
     & (K)  &  (cm\,s$^{-2}$) &   & (Gyr)  & (M$_{\odot}$)&  (\kms) & \\
\noalign{\smallskip}
\hline    
\noalign{\smallskip}
 \object{HD\, 142860}  &    6361  &   4.07 &  -0.09  &  2.89  &  1.21  &   10.65 &        2.32   \\ 
 \object{HD\, 89569 }  &    6469  &   4.08 &   0.09  &  2.14  &  1.41  &   11.33 &   $<$  1.90   \\
 \object{HD\, 86264 }  &    6300  &   4.06 &   0.25  &  2.19  &  1.39  &   12.55 &        1.50   \\ 
 \object{HD\, 210302}  &    6405  &   4.24 &   0.10  &  1.50  &  1.29  &   13.68 &        2.42   \\ 
 \object{WASP-3     }  &    6423  &   4.42 &   0.04  &  0.93  &  1.27  &   15.21 &        2.60   \\ 
 \object{HD\, 30652 }  &    6494  &   4.29 &   0.04  &  0.84  &  1.29  &   17.01 &        2.25   \\ 
 \object{30AriB     }  &    6284  &   4.35 &   0.12  &  2.55  &  1.32  &   42.61 &        2.72   \\ 
 \object{HAT-P-41   }  &    6479  &   4.39 &   0.13  &  1.05  &  1.35  &   20.11 &   $<$  1.70   \\ 
 \object{HAT-P-2    }  &    6414  &   4.18 &   0.04  &  1.91  &  1.30  &   20.50 &   $<$  1.00   \\ 
 \object{HAT-P-34   }  &    6509  &   4.24 &   0.08  &  1.17  &  1.35  &   24.08 &   $<$  1.50   \\ 
 \object{HD\, 8673  }  &    6472  &   4.27 &   0.14  &  0.55  &  1.24  &   26.91 &        1.50   \\ 
 \object{CoRoT-11   }  &    6343  &   4.27 &   0.04  &  1.58  &  1.28  &   36.72 &        2.10   \\ 
\noalign{\smallskip} 
\hline
\hline
\end{tabular}
\label{tabla_maria}
\end{table*}

\begin{figure}
\centering
\includegraphics[width=9cm]{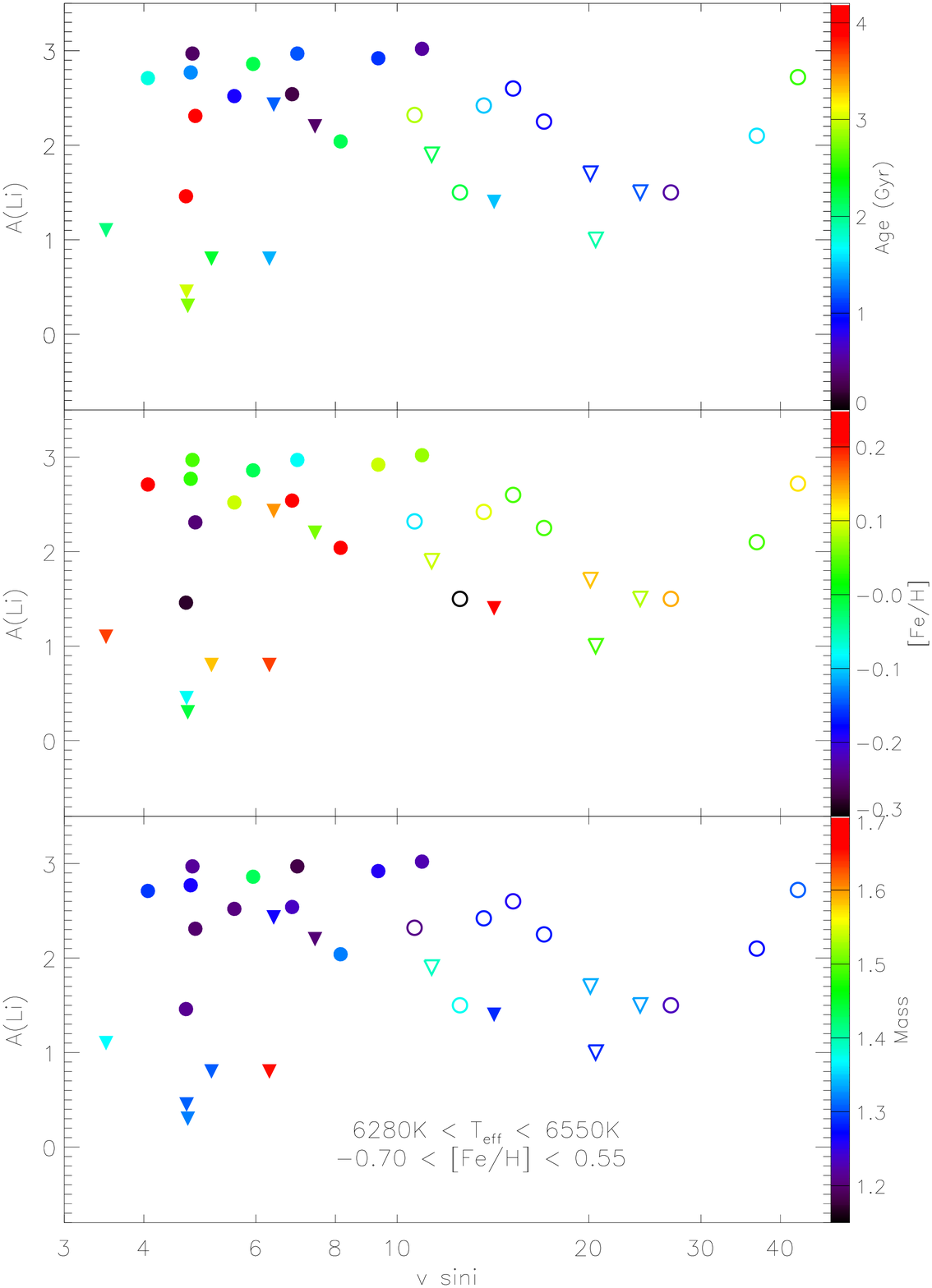}
\caption{Lithium abundances vs. v \textit{sin}i around the Li dip for all our stars (filled circles and triangles). Open circles and triangles (upper limits) represent the fast rotators from \citet{tsantaki14}. In each panel a color scale shows the ages, metallicities and masses of the stars.} 
\label{Li dip}
\end{figure}

Several mechanisms have been proposed to explain the formation of the Li dip, such as mass loss \citep{schramm90}, diffusion and radiative acceleration \citep{richer93} or rotationally induced mixing \citep{zahn92,pinsonneault90}. Under the assumption of this last mechanism, stars that rotate faster in the MS will experience more rapid mixing (thus, more Li depletion) than slow rotators at the same mass \citep{pinsonneault97}. In principle, this seems to be at odds with the work by \citet{bouvier08} (where the slow rotators on the ZAMS suffer a stronger depletion of Li than fast rotators) and with the higher Li abundances found in rapidly rotating stars as compared with slow rotators of the same mass in the Pleiades \citep{soderblom93,garcialopez} or IC 2602 \citep{randich97}. However, at this point one has to be careful distinguising between the depletion mechanisms acting during the PMS and the MS \citep[e.g.][]{somers14} and between the rotation rates in the ZAMS and the current rotation rates, though a priori one could expect that a star with a current high velocity was also a fast rotator in the past. Nevertheless, it is difficult to have an estimation of the initial rotation velocity since stars usually spin down when arriving at the MS. Therefore we have to extract information from the current surface rotation rates. Moreover, for most of the stars we do not know their inclination so we can only measure the projected rotational velocity, v \textit{sin}i, given that we have a good quality spectrum.\\

In order to check for the possible impact of rotation we compare the v \textit{sin}i values of the stars that typically form the Li gap (which we define in the range 6280 K $<$ \teff\ $<$ 6550 K, see Fig. \ref{Li dip}). We note that in this plot the number of stars is lower since we were able to determine v \textit{sin}i for only 20 out of 50 stars in this \teff\ range. To increase the statistics we included 12 fast rotators within the same \teff\ range analysed in \citet{tsantaki14} (see Table \ref{tabla_maria}). For those stars the stellar parameters are derived using the spectral synthesis technique for FGK stars and are in agreement with the results of the EW method. In addition, the comparison of v \textit{sin}i from the spectral synthesis technique mentioned above and our method shows a good agreement. The Li abundances for those 12 objects were derived in the same way as in this work. Therefore, the addition of the extra stars in Fig. \ref{Li dip} guarantees a uniform comparison.\\

We can observe that the stars with the lowest Li abundances ($<$ 1 dex) show low v \textit{sin}i values (3-6 \kms) whereas the stars with the highest Li, which presumably surround the dip, present a wide range of rotation rates ($\sim$4-10 \kms). This could be caused by the fact that the dip stars are older on average than the stars with higher Li at the same v \textit{sin}i (upper panel of Fig. \ref{Li dip}). There seems to be a slight increase of the upper envelope of Li abundances with rotation up to 10 \kms, however, from that point Li abundances decrease sharply as the rotation increases. The fast rotators that form the upper envelope of Li abundances (within 10-30 \kms) have similar ages ($<$ 1.5 Gyr) than the slower stars with high Li content. In the lower panels of Fig. \ref{Li dip} it can be seen that the metallicities and masses of both groups are comparable, thus we could expect that the higher rotation is producing extra depletion in these stars, as suggested by rotationally-induced mixing models. We could think then that at young ages ($<$ 1.5 Gyr) the Li dip is only formed by fast rotators while for older ages the stars have had more time to deplete Li and to spin down at the same time, making it impossible to distinguish between rotationally induced mixing and other destruction mechanisms acting during the MS. \\

We checked if the lack of stars with v \textit{sin}i determination and Li upper limits could be biasing this effect. We could not determine v \textit{sin}i values for 9 stars (with Li upper limit) but only one out of those has a young age (WASP-32, 0.7 Gyr) and seems a slow rotator by observing its spectrum. Thus, this is the only young star in our sample belonging to the dip with a low rotation rate. In any case, the small number of stars in this subsample suggests to take this result with caution. Moreover, the rotational models predict a correlation between the rotation history of the star and Li depletion, rather than a correlation between the current rotation and the Li abundance \citep{pinsonneault97}. There is another group of 5 fast rotators within the same v \textit{sin}i range (10-30 \kms) forming a lower envelope for Li abundances, probably related to their greater ages (though we note that these are the only stars with \logg\ $<$ 4.2 in Figure \ref{Li dip} and we cannot rule out the possibility that they are subgiants). Curiously, for the two objects with the highest v \textit{sin}i the trend changes, showing a higher Li content despite being older. We should consider with caution this rise in Li abundance since the determination of parameters becomes more difficult for the fastest rotators and the errors are three times larger than for the non-rotating counterparts.\\

\subsection{Li evolution: dependence on [Fe/H] and age}\label{li-feh}

To extract information about the evolution of Li through the life of the galaxy it is very common to evaluate its behaviour with the metallicity. The well known 'Spite plateau' shows how the abundances of Li are nearly constant at [Fe/H] $\lesssim$ -1 dex while they increase as [Fe/H] increases. However, the available studies of clusters and field stars do not include very metal-rich stars with the exception of the recently analyzed cluster NGC6253 \citep{cummings12} with [Fe/H] = 0.43 dex. We note that a quite lower metallicity ([Fe/H] = 0.23 dex) has been obtained for this cluster by other authors \citep{montalto12}. Our sample of metal-rich planet hosts represents a good opportunity to check how the Li abundances behave at [Fe/H] $>$ 0.2 dex.\\

\begin{figure}
\centering
\includegraphics[width=9.0cm]{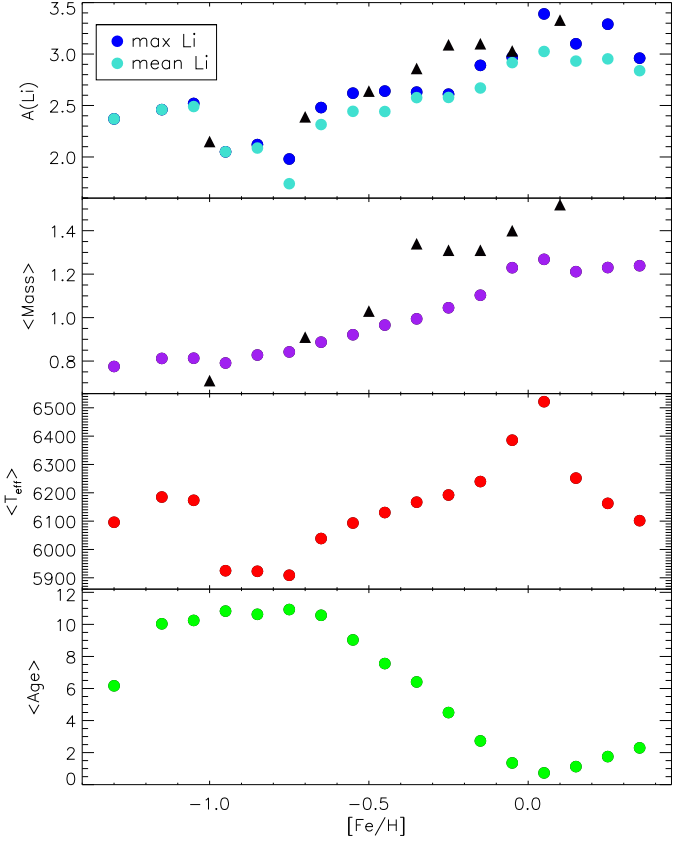}
\caption{\textit{Upper panel:} Maximum and mean values of Li in different metallicity bins for the 6 stars with the highest Li abundance in each metallicity bin (with \logg$>$4.2). The circles are the values from this work and the triangles denote the values from \citet{lambert04}. \textit{Rest of the panels:} Mean values of mass, \teff\ and age for those 6 stars in each metallicity bin.} 
\label{Li-evol}
\end{figure}

In Fig. \ref{Li-evol} we show the mean values of Li, stellar mass, \teff\ and age for the six stars with the highest Li abundance in each bin of metallicity. We chose this number of stars per bin in order to compare our results with the values obtained by \citet{lambert04} in a similar approach, who in turn, reported a good agreement with the maximum values found in open clusters of similar metallicity. We should consider with caution the most metal-poor bins ([Fe/H] $<$ -0.7 dex) since we only have one or two stars per bin and their temperatures fall out of the main trend. We compared our parameters with the ones derived by \citet{casagrande11} for these metal-poor stars and they agree well except for the most metal-poor star (HD31128) for which \citet{casagrande11} gives a higher age, 8.26 Gyr.  \\

\begin{figure}
\centering
\includegraphics[width=9.0cm]{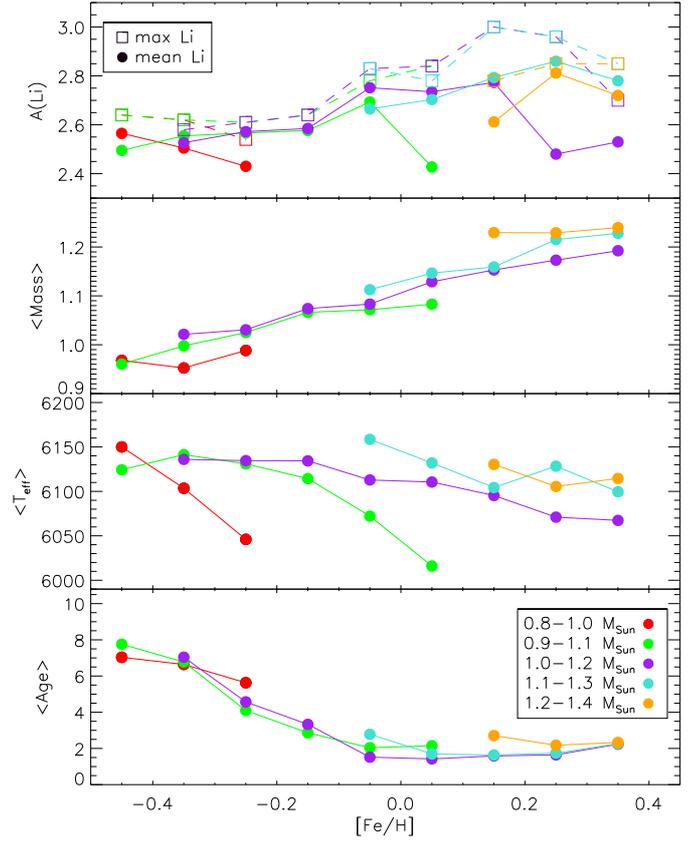}
\caption{\textit{Upper panel:} Maximum and mean values of Li in different metallicity bins for the 6 stars (when available) with the highest Li abundance in each metallicity bin (with 6000K $<$ \teff\ $<$ 6200K and \logg$>$4.2). \textit{Rest of the panels:} Mean values of mass, \teff\ and age for those 6 stars in each metallicity bin. } 
\label{Li-evol-teff}
\end{figure}

For metallicities lower than solar the Li abundance increases steadily with metallicity from a minimum at [Fe/H] = -0.75 dex with A(Li)$\sim$2, close to the 'Spite plateau', till [Fe/H] $\sim$ 0 dex. What is the reason of the abundances increasing from there till the solar metallicity? First, we can expect the more metallic stars to retain more of their initial Li because their masses are increasing (and thus convective envelopes becoming shallower). Second, metal-rich stars are younger so they have less time to deplete their Li as observed in younger clusters \citep[e.g.][]{sestito05}. Finally, the models of Li production point to an increase of Li with time in the interstellar medium \citep[e.g.][]{fields99,prantzos12}.\\

On the other hand, for metallicities higher than solar, Li abundances seem to flatten and even decrease for the most metal-rich stars. The same happens to the age, as expected, it decreases as [Fe/H] increases to reach a plateau for the most Li rich stars, where the age is between 1-2 Gyr. The maximum Li abundance, A(Li)=3.39, is found at solar metallicity which coincides with the minimum in age. We note here that this value corresponds to WASP-66, a very young and quite hot star (\teff\ = 7051K) compared with the average \teff\ in our sample, hence the rise in average \teff\ showed in the third panel of Fig. \ref{Li-evol}. That value also matches the maximum Li abundances found in young clusters such as NGC2264 \citep{sestito05} and is very similar to the meteoritic abundance. Therefore it is possible that this represents the initial maximum Li abundance and those stars have not experienced any astration.\\

\begin{figure*}
\centering
\includegraphics[width=13.5cm]{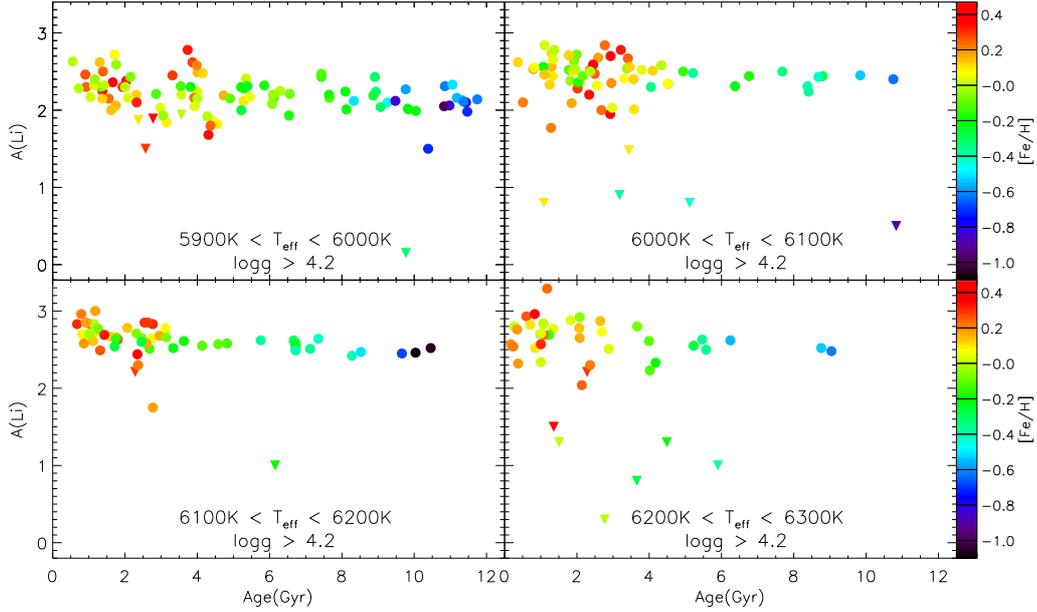}
\caption{Li abundances a  function of age in several \teff\ regions. The metallicity values are shown with a colour scale.} 
\label{Li-age}
\end{figure*}

The standard model predicts that Li depletion is faster for more metallic stars since they have deeper convective zones. This is in contrast with the higher Li abundances found for the metal-rich stars. However, the models of Galactic Li production predict that initial Li abundance in a star becomes higher as the Galaxy evolves, i. e., as [Fe/H] increases. Indeed, the high Li abundances found in meteorites or in young clusters require Galactic production to increase the primordial Li abundance (either $\sim$2.2, from 'Spite plateau' or $\sim$2.7 from WMAP observations). Maybe, the flattening observed at high [Fe/H] is a balance between the higher initial Li in the stars (due to a higher content of Li in the interstellar medium) and a stronger destruction of it due to the deepening of stellar convective zones. This possibility is suggested by the models of \citet{fields99} who show that at super-solar metallicities the stellar Li depletion begins to affect the Li abundance in the inter-stellar medium, and thus flattening the correlation of initial Li and Fe. Therefore, in this scenario, we may think that the stars around solar metallicity represent the maximum Li in the Galaxy, which is similar to the initial value (and to the meteoritic value) since they are young and have not depleted it yet. As you move to super solar metallicities the effect reverses and the high [Fe/H] begins to produce Li depletion. However, we should also note in Fig. \ref{Li-evol} that the \teff\ is also decreasing as [Fe/H] increases, therefore the convective envelope is getting deeper\footnote{\citet{pinsonneault01} show that \teff\ is the main parameter determining the mass of convective envelopes with a very small effect of metallicity} and probably affecting the depletion of Li. \\

The question remains how we can distinguish between a lower content of Li due to a lower initial abundance ([Fe/H] effect on Galactic production) or due to internal destruction during the MS (\teff\ effect). In an atempt to disentangle both effects we constructed samples of stars with different mass ranges but with 6000K $<$ \teff\ $<$ 6200K as shown in Fig. \ref{Li-evol-teff}. We chose this \teff\ range because it is well populated with stars of different masses and metallicities. As expected, we cannot observe the stars with the highest Li content since we are not using the hottest stars. By restricting the sample, now the average \teff, mass, and age of the stars are very similar in all the bins at super-solar metallicities, thus the observed variation of Li abundances should be triggered basically by the metallicity variation. We still observe the increase of Li with [Fe/H], with a maximum at [Fe/H] = 0.15 dex and a clear decrease for the subsamples between 1M$_{\odot}$ and 1.3M$_{\odot}$ (the most populated ones, purple and blue symbols). We evaluated the possible effect of planets on Li evolution, since as suggested before, planet hosts seem to have depleted more Li and our most metal-rich bins contain many of them. Thus, we reconstructed the samples between 1M$_{\odot}$ and 1.3M$_{\odot}$ with only comparison stars and we found that the behaviour is similar, i.e., Li decreases for the most metal-rich stars from [Fe/H] $\sim$ 0.15 dex. Therefore, the lowest Li abundances found in the most metal-rich stars seem to be caused by a lower initial Li as predicted by some models of galactic production \citep{fields99}. \\

It is commonly accepted that Li abundances decrease with age though the main depletion takes place principally during very young ages and depends on initial rotation rates \citep[e.g.][]{charbonnel05} whereas after 1-2 Gyr the age effect is not so strong \citep{sestito05}. In Fig. \ref{Li-age} we show Li abundance as a function of age in four different \teff\ bins. It is quite clear that the higher abundances appear in the younger objects and then the Li upper envelope slightly decreases to reach a kind of a plateau. However we can still observe a high dispersion in Li for stars with similar \teff, metallicity and age. For example, in the top-right panel, (6000K-6100K), for the most metallic stars ([Fe/H] $\sim$ 0.4 dex, red symbols) there is a dispersion of $\sim$0.8 dex in Li abundance determinations (not considering upper limits). For stars of solar metallicity (green points) the dispersion reaches 0.4 dex for stars of similar age. We find a similar spread in other \teff\ and [Fe/H] regions and it can reach values of 2 dex if we consider the upper limits. 
This fact reveals that an extra parameter is governing Li depletion. We may note here that when dealing with MS stars, the age determination is probably very uncertain \citep[e.g.][]{jorgensen2}, at least significantly more uncertain than other stellar parameters determination. Therefore, it might be possible that these stars of apparently same age could have quite different ages and that could be the reason of the spread in abundances. For instance, in clusters like the Hyades or NGC6243 the spread around 6000K is quite small \citep{cummings12}. On the other hand, M67 or Praesepe show a huge dispersion in Li abundances \citep{sestito05}.  \\

\subsection{Li in the galactic disks}\label{li-galaxy}
In the recent work by \citet{ramirez_li12} a first attempt to compare Li abundances in the two galactic disks was presented. They found that the Li abundances for the thin disk stars (using a kinematical separation criteria) increase with metallicity 
while for the thick disk stars the abundances have a nearly constant maximum value of A(Li)$\sim$2.1 dex, similar to the 'Spite plateau'. However, once they cleaned their thin disk sample of the youngest and more massive stars (to allow a 
less biased comparison with the older and cooler thick disk) they found a smoother transition from the thick to the thin 
disk.\\

\begin{figure}
\centering
\includegraphics[width=9.0cm]{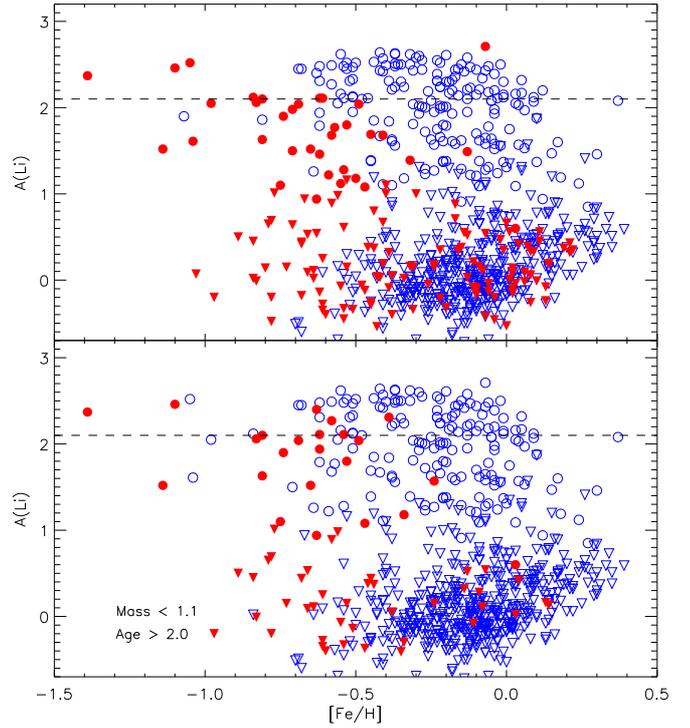}
\caption{Li abundances as a function of metallicity for the HARPS non-host stars. Thin disk stars are depicted with blue symbols while thick disk stars are denoted with red symbols. Downward triangles are upper limits on Li. The separation of the galactic populations are based on the abundances (\textit {top}) and kinematics (\textit {bottom}). The constrain on age is only applied for thin disk stars as in \citet{ramirez_li12}.} 
\label{Li-disk}
\end{figure}

In Fig \ref{Li-disk} we present a plot similar to Fig 5. of \citet{ramirez_li12} using only the stars of the HARPS samples analyzed in \citet{adibekyan12c}. We note that we have removed planet hosts for this section since they represent only $\sim$10\% of the total sample and their abundances might be affected by the presence of planets (at least in the solar \teff\ region). In order to allow a better comparison we add a set of cool stars (\teff\ $<$ 5600K) belonging to the HARPS samples (see Table 6). We used both kinematic\footnote{The kinematic separation was done using the prescription of \citet{bensby03} as presented in \citet{adibekyan12c}.} and chemical criteria to separate the stellar populations \citep[see][for details]{adibekyan11, adibekyan12c}. We note that stars with [Fe/H] $>$ -0.2 dex and showing enhancement in $\alpha$-element abundances were classified as members of a high-$\alpha$ metal-rich population in \citet{adibekyan11, adibekyan13}. Here we use the same symbol as for the thick disk stars to compare easier with the results of \citet{ramirez_li12}. \\
 
Our thin disk stars also show slightly higher maximum abundances, decreasing for the more metal-rich stars as in the above mentioned work and reflecting the evolution of Li at high metallicities discussed in the previous section. However, our thick disk stars also show a decrease of Li with metallicity from [Fe/H] $>$ -0.5 dex, whereas in \citet{ramirez_li12} the thick disk stars present a constant value close to the 'Spite plateau' till [Fe/H] $\sim$ -0.1 dex. Furthermore, this decrease seems steeper for thick disk stars than for the thin disk. The lack of Li-rich metal-rich thick disk stars in our sample when compared to that observed in \citet{ramirez_li12} can probably be explained by the different criteria used to separate the stellar populations. However, both our kinematic and chemical separation shows the same picture. We should note that our kinematic criteria suggest very few thick disk stars with [Fe/H] $>$ -0.3 dex, while in the sample of \citet{ramirez_li12} this metallicity region is quite abundant of thick disk stars. A more detailed analysis of their metal-rich Li-rich thick disk stars is needed to understand the nature of these stars and the reason of the observed discrepancy.\\

\section{Summary}

We present new Li abundances for a total sample of 36 planet hosts and 229 stars without detected giant planets in the HARPS GTO samples, together with 88 additional extrasolar planet hosts from other sources. All these stars span over an effective temperature range 5900\,K\,$<$\,\teff\,$<$\,7200\,K. First, we find that planet hosts show an extra depletion of 0.07 dex in the \teff\ range 5900-6300K as previouly claimed by \citet{gonzalez_li15}. This offset is statistically significant but close to the average uncertainties of Li abundances. However, this offset seem to be stronger for stars hosting hot jupiters (0.14 dex), something that could be explained by some models where the effect of planets on Li depletion is related with their mass and migration. This issue should be explored in the future with a larger sample of hot jupiters than the currently analyzed here (24 stars). On the other hand, if we include the Li dependence on v \textit{sin}i in our multivariate regression fit, the offset in Li abundance between the planet hosts and the comparison stars becomes positive but also decreases and becomes insignificant. This suggest that the difference in v \textit{sin}i between both samples was causing the difference in Li abundances. Nevertheless, the number of stars for which we can derive v \textit{sin}i is still very low to achieve a conclusion regarding the effect of rotation on Li abundances for our sample of planet hosts.\\

We study the position of stellar mass of the Li dip at several metallicity bins. We confirm that the mass of the Li dip increases with the metallicity and extend this relation till [Fe/H] = 0.4 dex. However, the mass of all the stars in our sample increase with metallicity, thus jreflecting the mass-metallicity relation for dwarfs of the same temperature and supporting the idea of a constant \teff\ for the Li dip. We also evaluate the behaviour of Li abundances with v \textit{sin}i for the stars that surround the dip. We find that for the younger objects ($\lesssim$ 1.5 Gyr), a strong depletion of Li only happens for fast rotators ($\gtrsim$ 10 \kms), suggesting that the Li dip is formed due to rotationally induced mixing at early stages of the MS. However, for the older objects the Li dip is formed by slower rotators, making it impossible to differentiate between the previously mentioned mechanism (we do not know if those stars were fast rotators at younger ages) or other depletion processes taking place during the MS.\\

Finally we analyze the Li evolution with the metallicity (i.e., the age of the Galaxy) since our metal-rich sample represents a good opportunity to check the behaviour of Li at super-solar metallicities, not so well studied in the literature. As expected from models of Galactic production of Li, we observe an increase of Li abundances as the Galaxy evolves, i.e. as the metallicity increases. We find the maximum abundance around [Fe/H] $\sim$ 0.1 dex, with A(Li) = 3.39 dex, which is similar to the meteoritic value and the maximum Li abundances found in young clusters such as NGC2264 \citep{sestito05}. On the other hand, Li abundances flatten and even decrease for the most metal-rich stars. This is in agreement with the models of \citet{fields99} which suggest that the initial Li abundance of the most metal-rich stars is lower because the interstellar Li abundances have decreased due to the impact of stellar depletion during the evolution of the Galaxy. We also study the behaviour of Li in the context of thin and thick disks. We find a clear decrease of Li abundances at super-solar metallicities for the thin disk and a steeper decrease for thick disk stars that starts at [Fe/H] $>$ -0.5 dex.\\

\begin{acknowledgements}
E.D.M, S.G.S and V.Zh.A. acknowledge the support from the Funda\c{c}\~ao para a Ci\^encia e Tecnologia, FCT (Portugal) in the form of the grants SFRH/BPD/76606/2011, SFRH/BPD/47611/2008 and SFRH/BPD/70574/2010, respectively. PF acknowledges support by Funda\c{c}\~ao para a Ci\^encia e a Tecnologia (Portugal) through Investigador FCT contracts of reference IF/01037/2013 and POPH/FSE (EC) by FEDER funding through the program ``Programa Operacional de Factores de Competitividade - COMPETE''. J.I.G.H. acknowledges support from the Spanish Ministry of Economy and Competitiveness (MINECO) under the 2011 Severo Ochoa Program MINECO SEV-2011-0187. NCS, A.M. and M.T. thanks for the support by the European Research Council/European Community under the FP7 through Starting Grant agreement number 239953, as well as the support through programme Ci\^encia\,2007 funded by FCT/MCTES (Portugal) and POPH/FSE (EC), and in the form of grant PTDC/CTE-AST/098528/2008. AM is supported by the European Union Seventh Framework Programme (FP7/2007-2013) through grant agreement number 313014 (ETAEARTH).

This research has made use of the SIMBAD database operated at CDS, Strasbourg (France) and the Encyclopaedia of extrasolar planets.\\
This work has also made use of the IRAF facility.

\end{acknowledgements}

\bibliographystyle{aa}
\bibliography{edm_bibliography}

\begin{table*}
 \caption{Li abundances for stars with planets from HARPS GTO samples. Parameters from
\citet{sousa08,sousa_harps4,sousa_harps2}}
\centering
\begin{tabular}{lcccrrcrccc}
\hline
\hline
\noalign{\smallskip}
Star & $T_\mathrm{eff}$ & $\log{g}$ & $\xi_t$ & [Fe/H] & Age & Mass & ${\rm A(Li)}$ & error & vsini & hot jupiter\\  
     & (K)  &  (cm\,s$^{-2}$) & (km\,s$^{-1}$) &  & (Gyr)  & (M$_{\odot}$) & & &  (\kms) & \\
\noalign{\smallskip}
\hline    
\noalign{\smallskip}
        \object{HD\, 142 }  &  6403  &  4.62  &  1.74  &   0.09  &     1.08  &     1.27  &      2.92  &   0.04  &     9.34  &    no   \\
      \object{HD\, 10647 }  &  6218  &  4.62  &  1.22  &   0.00  &     0.26  &     1.16  &      2.80  &   0.03  &     5.30  &    no   \\
      \object{HD\, 17051 }  &  6227  &  4.53  &  1.29  &   0.19  &     0.17  &     1.21  &      2.57  &   0.05  &     5.56  &    no   \\
      \object{HD\, 19994 }  &  6289  &  4.48  &  1.72  &   0.24  &     2.14  &     1.34  &      2.04  &   0.07  &     8.15  &    no   \\
      \object{HD\, 23079 }  &  5980  &  4.48  &  1.12  &  -0.12  &     6.12  &     1.00  &      2.16  &   0.04  &     2.96  &    no   \\
      \object{HD\, 39091 }  &  6003  &  4.42  &  1.12  &   0.09  &     2.07  &     1.11  &      2.34  &   0.04  &     2.96  &    no   \\
      \object{HD\, 52265 }  &  6136  &  4.36  &  1.32  &   0.21  &     1.04  &     1.20  &      2.83  &   0.05  &     ----  &    no   \\
      \object{HD\, 75289 }  &  6161  &  4.37  &  1.29  &   0.30  &     0.68  &     1.21  &      2.83  &   0.04  &     4.30  &    no   \\
      \object{HD\, 82943 }  &  5989  &  4.43  &  1.10  &   0.26  &     0.90  &     1.15  &      2.46  &   0.04  &     2.75  &    no   \\
     \object{HD\, 108147 }  &  6260  &  4.47  &  1.30  &   0.18  &     0.37  &     1.23  &      2.32  &   0.03  &     5.85  &    no   \\
     \object{HD\, 117618 }  &  5990  &  4.41  &  1.13  &   0.03  &     4.00  &     1.08  &      2.24  &   0.03  &     3.67  &    no   \\
     \object{HD\, 121504 }  &  6022  &  4.49  &  1.12  &   0.14  &     1.30  &     1.14  &      2.56  &   0.03  &     3.61  &    no   \\
     \object{HD\, 169830 }  &  6361  &  4.21  &  1.56  &   0.18  &     2.04  &     1.39  &  $<$ 1.10  &    --   &     3.49  &    no   \\
     \object{HD\, 179949 }  &  6287  &  4.54  &  1.36  &   0.21  &     0.24  &     1.24  &      2.54  &   0.04  &     6.84  &   yes   \\
     \object{HD\, 196050 }  &  5917  &  4.32  &  1.21  &   0.23  &     3.91  &     1.12  &      2.16  &   0.03  &     3.34  &    no   \\
     \object{HD\, 208487 }  &  6146  &  4.48  &  1.24  &   0.08  &     0.82  &     1.17  &      2.70  &   0.04  &     4.01  &    no   \\
     \object{HD\, 209458 }  &  6118  &  4.50  &  1.21  &   0.03  &     1.19  &     1.13  &      2.73  &   0.05  &     ----  &   yes   \\
     \object{HD\, 212301 }  &  6271  &  4.55  &  1.29  &   0.18  &     0.35  &     1.24  &      2.76  &   0.04  &     5.76  &   yes   \\
     \object{HD\, 213240 }  &  5982  &  4.27  &  1.25  &   0.14  &     4.01  &     1.19  &      2.49  &   0.05  &     3.50  &    no   \\
     \object{HD\, 216435 }  &  6008  &  4.20  &  1.34  &   0.24  &     3.41  &     1.28  &      2.67  &   0.04  &     5.13  &    no   \\
     \object{HD\, 221287 }  &  6374  &  4.62  &  1.29  &   0.04  &     0.33  &     1.22  &      2.97  &   0.04  &     4.77  &    no   \\
       \object{HD\, 7449 }  &  6024  &  4.51  &  1.11  &  -0.11  &     1.77  &     1.06  &      2.52  &   0.03  &     3.51  &    no   \\
      \object{HD\, 10180 }  &  5911  &  4.39  &  1.11  &   0.08  &     4.55  &     1.06  &      1.82  &   0.03  &     2.80  &    no   \\
      \object{HD\, 93385 }  &  5977  &  4.42  &  1.14  &   0.02  &     3.56  &     1.07  &      2.20  &   0.03  &     2.99  &    no   \\
     \object{HD\, 134060 }  &  5966  &  4.43  &  1.10  &   0.14  &     1.75  &     1.12  &      2.06  &   0.04  &     3.21  &    no   \\
\noalign{\smallskip}							      
\hline
\noalign{\smallskip}
HARPS-4  & & & & & & & & &\\
\noalign{\smallskip}
\hline    
\noalign{\smallskip}							      
     \object{HD\, 190984 }  &  6007  &  4.02  &  1.58  &  -0.49  &     4.60  &     1.16  &  $<$ 0.50  &    --   &     3.23  &    no   \\
\noalign{\smallskip}							      
\hline
\noalign{\smallskip}
HARPS-2  & & & & & & & & &\\
\noalign{\smallskip}
\hline    
\noalign{\smallskip}							      
     \object{HD\, 125612 }  &  5913  &  4.43  &  1.02  &   0.24  &     1.39  &     1.10  &      2.50  &   0.05  &     ----  &    no   \\
     \object{HD\, 145377 }  &  6054  &  4.53  &  1.11  &   0.12  &     1.25  &     1.12  &      2.33  &   0.04  &     3.76  &    no   \\
     \object{HD\, 148156 }  &  6251  &  4.51  &  1.36  &   0.25  &     0.60  &     1.21  &      2.93  &   0.02  &     5.41  &    no   \\
     \object{HD\, 153950 }  &  6074  &  4.39  &  1.23  &  -0.01  &     4.34  &     1.12  &      2.58  &   0.04  &     3.41  &    no   \\
     \object{HD\, 156411 }  &  5910  &  3.99  &  1.31  &  -0.11  &     4.21  &     1.25  &  $<$ 0.30  &    --   &     3.34  &    no   \\
     \object{HD\, 217786 }  &  5966  &  4.35  &  1.12  &  -0.14  &     7.65  &     1.02  &      2.16  &   0.07  &     ----  &    no   \\
      \object{HD\, 25171 }  &  6160  &  4.43  &  1.22  &  -0.11  &     4.13  &     1.09  &      2.55  &   0.06  &     ----  &    no   \\
      \object{HD\, 72659 }  &  5926  &  4.24  &  1.13  &  -0.02  &     6.29  &     1.06  &      2.25  &   0.07  &     ----  &    no   \\
       \object{HD\, 8535 }  &  6158  &  4.42  &  1.25  &   0.04  &     1.76  &     1.15  &      2.65  &   0.03  &     3.07  &    no   \\
       \object{HD\, 9578 }  &  6055  &  4.52  &  1.07  &   0.11  &     1.30  &     1.12  &      2.74  &   0.04  &     2.36  &    no   \\
\noalign{\smallskip}
\hline
\noalign{\smallskip}
\end{tabular}
\label{tabla_harps_plan}
\end{table*}

\end{document}